\documentclass[12pt]{iopart}

\usepackage{iopams}

\usepackage{graphicx}
\usepackage{chemarrow}
\usepackage{amssymb}
\usepackage{color}
\begin{document}

\title[Projection-based filtering for stochastic reaction networks]
{Projection-based filtering for stochastic reaction networks}

\author{Shinsuke Koyama}

\address{Department of Statistical Modeling, The Institute of Statistical Mathematics, Tokyo, Japan}
\ead{skoyama@ism.ac.jp}
\vspace{10pt}
\begin{indented}
\item[]February 2016
\end{indented}

\begin{abstract} 
This study concerns online inference (i.e., filtering) on the state of reaction networks, conditioned on noisy and partial measurements. The difficulty in deriving the equation that the conditional probability distribution of the state satisfies stems from the fact that the master equation, which governs the evolution of the reaction networks, is analytically intractable. 
The linear noise approximation (LNA) technique, which is widely used in the analysis of reaction networks, has recently been applied to develop approximate inference. Here, we apply the projection method to derive approximate filters, and compare them to a filter based on the LNA numerically in their filtering performance. We also contrast the projection method with moment-closure techniques in terms of approximating the evolution of stochastic reaction networks.

\end{abstract}

% Uncomment for PACS numbers
%\pacs{00.00, 20.00, 42.10}
%
% Uncomment for keywords
%\vspace{2pc}
%\noindent{\it Keywords}: XXXXXX, YYYYYYYY, ZZZZZZZZZ
%
% Uncomment for Submitted to journal title message
%\submitto{\JPA}
%
% Uncomment if a separate title page is required
%\maketitle
% 
% For two-column output uncomment the next line and choose [10pt] rather than [12pt] in the \documentclass declaration
%\ioptwocol
%

%%%%%%%%%%%%%%%%%%%%%%%%%%%%%%%%%%%%%%%%%%%
\section{Introduction}
%%%%%%%%%%%%%%%%%%%%%%%%%%%%%%%%%%%%%%%%%%%

Stochastic reaction networks provide probabilistic descriptions of the evolution of interacting species. They are used for modeling phenomena in a wide range of disciplines; those species can represent molecules in chemical reactions \cite{Higham08,Thattai01,Shahrezaei08}, animal species in ecology \cite{Spencer05}, susceptibles and infectives in epidemic models \cite{Pastor-Satorras15}, and information packets in telecommunication networks \cite{Adas97}. 

The evolution of a network is modeled by a continuous-time Markov jump process, for which the probability distribution of the number of individuals of each species obeys the master equation \cite{Gardiner85,Kampen92}. 
Here, we consider a situation wherein only noisy and partial measurements of underlying reaction networks are available. 
Our objective is to infer the number of individuals of species from the observations obtained up to the current time. In the literature on signal processing, this problem is called {\it filtering} \cite{Jazwinski70}.

The filtering equation, which governs the posterior distribution conditioned on the observations, is not analytically obtainable due to the intractability of the master equation. 
It is possible to perform exact numerical simulation and obtain samples from the Markov jump processes using a stochastic simulation algorithm (SSA) \cite{Gillespie07}.  
Simulating many ``particles" with the SSA and sampling the weighted particles in the favor of the observations, we could obtain samples from the posterior distribution. This technique is known as the {\it sequential Monte Carlo method} or {\it particle filtering} \cite{Doucet01}.
However, the SSA is often too slow. Moreover, particle filtering sufficiently requires many particles to obtain precise posterior expectations. 
Thus, particle filtering might not be efficient for performing online inference. 

An alternative approach is to consider the suitable approximations of the Markov jump processes. 
In the linear noise approximation (LNA), 
which is most widely used in such analysis, 
a Gaussian process whose mean obeys the deterministic rate equation approximates a Markov jump process  \cite{Kampen92}. 
The LNA is valid under the assumption that the number of individuals of a species is large \cite{Kurtz71}. 
It is also exact for all systems with affine propensities as well as for some systems with nonlinear propensities \cite{Grima15}.
As the Gaussian process is tractable, The LNA allows us to derive an analytical expression of the approximate filtering equation \cite{Fearnhead14}.
In addition to the LNA, 
a number of approximation techniques have been proposed such as system-size expansions \cite{Kampen92}, moment-closure approximations \cite{Grima12} and conditional moment equations \cite{Hasenauer14}, and have been applied to inference of model parameters \cite{Milner13,Frohlich16}.

In this study, we propose applying the projection method \cite{Brigo98,Brigo99} to 
derive approximate filters.  
In this method, the evolution of the probability distributions is constrained on a finite-dimensional family of densities through orthogonal projection onto the tangent space with respect to the Fisher metric. 
We derive the projection-based filter for stochastic reaction networks, and compare it to an approximate filter based on the LNA numerically in their filtering performance. 
We also contrast between the projection method and moment-closure techniques in terms of approximating the master equation.

%%%%%%%%%%%%%%%%%%%%%%%%%%%%%%%%%%%%%%%%%%%
\section{Method}
%%%%%%%%%%%%%%%%%%%%%%%%%%%%%%%%%%%%%%%%%%%
%%%%%%%%%%%%%%%%%%%%%%%%%%%%%%%%%%%%%%%%%%%
\subsection{Reaction networks}
%%%%%%%%%%%%%%%%%%%%%%%%%%%%%%%%%%%%%%%%%%%

Throughout the study, the transpose of a matrix $B$ is written $B^T$.
Let $\mathcal{X}=\{X_1,\ldots,X_n\}$ be $n$ species, and consider $m$ reactions among these species described by
\begin{equation}
\sum_{i=1}^n \nu_{ij}^{-}X_i  \autorightarrow{$k_j$}{} \sum_{i=1}^n \nu_{ij}^{+}X_i
\qquad j=1,\ldots,m
\label{eq:reacnet}
\end{equation}
where $\nu_{ij}^{-}$ and $\nu_{ij}^{+}$ are stoichiometric coefficients of reactants and products, respectively, and $k_j$ is the reaction rate constant. 
We denote by  $x=(x_1,\ldots,x_n)^T$  the discrete composition vector whose $i$th component, $x_i$, is the number of individuals of species $X_i$. 
Let $A=(\Delta x_{ij})$ be an $n\times m$ matrix, called the {\it net effect matrix}, whose $(i,j)$ element, $\Delta x_{ij} = \nu^+_{ij}-\nu^-_{ij}$, is the change in the number of individuals of the $i$th species after one step of the $j$th reaction. 
Let $h(x)=(h_1(x),\ldots,h_m(x))^T$ be the vector whose $j$th component, $h_j(x)$, is the rate of $j$th reaction, given as 
\begin{equation}
h_j(x) = k_j \prod_{i=1}^n
\Bigg(
\begin{array}{c}
x_i \\
\nu^-_{ij}
\end{array}
\Bigg)~.
\end{equation}
From the Markov property, it follows that the probability distribution over $x$ at time $t$, $P(x,t)$, is governed by the master equation \cite{Gillespie92,Gadgil05}: 
\begin{equation}
\frac{dP(x,t)}{dt} = \sum_{j=1}^m h_j(x-\Delta x_{\cdot j})P(x-\Delta x_{\cdot j},t)
- \sum_{j=1}^m h_j(x)P(x,t).
\label{eq:stochastic}
\end{equation}

Stochastic processes described by Eq.~(\ref{eq:stochastic}) are related to an ordinary differential equation (ODE), called the {\it rate equation}, via the thermodynamic limit. 
To see this, we introduce a scale factor $\Omega$ (typically taken to be ``volume"), 
and rescale the composition vector and the reaction rate as
\begin{equation}
z = \frac{x}{\Omega},
\label{eq:concentration}
\end{equation}
\begin{equation}
\tilde{h}(z) = \frac{h(\Omega z)}{\Omega}.
\label{eq:rate}
\end{equation}
Accordingly, the reaction rate constants are rescaled as 
\begin{equation}
\tilde{k}_j = V^{\sum_{i=1}^n \nu^-_{ij}-1} k_j
\qquad j=1,\ldots,m.
\end{equation} 
With these rescaled parameters, it has been proved in  \cite{Kurtz70} that $z\to\phi$ as $\Omega\to\infty$ in probability, where $\phi$ satisfies the rate equation:
\begin{equation}
\frac{d\phi}{dt} = A\tilde{h}(\phi). 
\label{eq:deterministic}
\end{equation}

%%%%%%%%%%%%%%%%%%%%%%%%%%%%%%%%%%%%%%%%%%%
\subsection{State space model and filtering}
%%%%%%%%%%%%%%%%%%%%%%%%%%%%%%%%%%%%%%%%%%%

We consider a situation wherein the system of interest is given by a stochastic reaction network, whose state is not directly observable, but instead, we have noisy and partial measurements at discrete time points \cite{Fearnhead14,Golightly06,Golightly11,Komorowski09,Finkenstadt13}; this situation is formulated within the framework of state space models. 
In state space modeling, the state process, $x(t)$, is given by the master equation (\ref{eq:stochastic}), and the measurement model is assumed to be
\begin{equation}
y_i = Gx(t_i) + \xi_i
\qquad i=1,\ldots N,
\label{eq:observation}
\end{equation}
where $y_i\in\mathbb{R}^d$ $(d\le n)$, $G\in\mathbb{R}^{d\times n}$, 
and $\xi_i$ is a  $d$-dimensional Gaussian random variable with zero mean and covariance matrix $V$. 
The goal of a filtering problem is to compute the posterior probability of the state $x$ at time $t_i$, when the observations $y_1,\ldots,y_i$ are given.

%%%%%%%%%%%%%%%%%%%%%%%%%%%%%%%%%%%%%%%%%%%
\subsection{Projection-based filter}
%%%%%%%%%%%%%%%%%%%%%%%%%%%%%%%%%%%%%%%%%%%
%%%%%%%%%%%%%%%%%%%%%%%%%%%%%%%%%%
\subsubsection{Projection method}

We apply the projection method proposed in \cite{Brigo98,Brigo99} to derive approximate filters.
To apply the projection method, we need a Fokker-Planck equation derived from the master equation (\ref{eq:stochastic}). 
By taking up to the second-order terms in the  Kramers-Moyal expansion of the master equation, a Fokker-Planck equation is obtained as 
\begin{eqnarray}
\frac{\partial p(x,t)}{\partial t} 
&=& 
\mathcal{L}^*p(x,t) \nonumber\\
&:=&
-\sum_{i=1}^n \frac{\partial}{\partial x_i}\Bigg[\sum_{k=1}^m\Delta x_{ik}h_k(x)p(x,t)\Bigg] 
\nonumber\\
& & { } + \frac{1}{2}\sum_{i,j=1}^n\frac{\partial^2}{\partial x_i\partial x_j}
\Bigg[
\sum_{k=1}^m \Delta x_{ik}h_k(x)\Delta x_{jk}
p(x,t)\Bigg]~,
\label{eq:Fokker-Planck}
\end{eqnarray}
where $p(x,t)$ is the probability density of $x$ at time $t$ \cite{Kampen92}. 
We apply the projection method to Eq.~(\ref{eq:Fokker-Planck}). 
The key idea is to introduce a finite-dimensional family of probability densities $p(x,\theta)$, 
where $\theta=(\theta_1,\ldots,\theta_r)\in \Theta \subseteq \mathbb{R}^r$ is the parameter characterizing the probability distributions, 
and to project the evolution of the probability density $p(x,t)$ onto the space of $p(x,\theta)$; the resulting ODE for $\theta$ approximates the master equation. 

Let $L_2$ be a space of square-integrable functions, and consider the square roots of the probability densities, $S^{1/2}=\{p(x,\theta)^{1/2}, \theta\in\Theta \} \subset L_2$. 
The tangent space of $S^{1/2}$ at $p(x,\theta)^{1/2}$ is given by
\begin{equation}
T_{p(x,\theta)^{1/2}} S^{1/2} = 
\mathrm{span}
\Bigg\{
\frac{\partial p(x,\theta)^{1/2}}{\partial\theta_1}, \ldots,
\frac{\partial p(x,\theta)^{1/2}}{\partial\theta_r}
\Bigg\}~.
\end{equation}
The $L_2$ inner product of any two bases of $S^{1/2}$ is defined as 
\begin{eqnarray}
\fl
\bigg\langle
\frac{\partial p(x,\theta)^{1/2}}{\partial\theta_i}, 
\frac{\partial p(x,\theta)^{1/2}}{\partial\theta_j}
\bigg\rangle &:=& 
\int
\frac{\partial p(x,\theta)^{1/2}}{\partial\theta_i}
\frac{\partial p(x,\theta)^{1/2}}{\partial\theta_j}
dx \nonumber\\
&=&
\frac{1}{4}\int \frac{\partial\log p(x,\theta)}{\partial\theta_i}
 \frac{\partial\log p(x,\theta)}{\partial\theta_j} p(x,\theta)dx \nonumber\\
&=&
 \frac{1}{4} g_{ij}(\theta),
\end{eqnarray}
where $(g_{ij}(\theta))$ is the Fisher information matrix.
Then, the orthogonal projection of $q\in L_2$ onto $T_{p(x,\theta)^{1/2}} S^{1/2}$ is given by 
\begin{equation}
q \mapsto 
\sum_{i=1}^r
\Bigg(
\sum_{j=1}^r 4g^{ij}(\theta)
\bigg\langle
q, \frac{\partial p(x,\theta)^{1/2}}{\partial\theta_j}
\bigg\rangle
\Bigg)
\frac{\partial p(x,\theta)^{1/2}}{\partial\theta_i} ~,
\label{eq:projection}
\end{equation}
where $(g^{ij})$ is the inverse of the Fisher information matrix.

Using Eq.~(\ref{eq:projection}), we project the Fokker-Plank equation (\ref{eq:Fokker-Planck})  onto $S^{1/2}$ as follows: 
Using the chain rule, we obtain the equation for $p(x,\theta)^{1/2}$ as
\begin{equation}
\label{eq:squareroot}
\frac{\partial p(x,\theta)^{1/2}}{\partial t} = \frac{p(x,\theta)^{1/2}\mathcal{L}^*p(x,\theta)}{2p(x,\theta)}.
\end{equation}
Applying the orthogonal projection (\ref{eq:projection}) to Eq.~(\ref{eq:squareroot}), we obtain an ODE for $\theta$ as 
\begin{equation}
\label{eq:projectionFP}
\frac{d\theta_i}{dt} = 
\sum_{j=1}^r g^{ij}(\theta) \mathrm{E}\Bigg[
\frac{\mathcal{L}^*p(x,\theta)}{p(x,\theta)}
\frac{\partial \log p(x,\theta)}{\partial \theta_j}
\Bigg]
\qquad  i=1,\ldots, r,
\end{equation}
where $\mathrm{E}[\cdot]$ is the expectation of $x(t)$ with respect to $p(x,\theta)$. 
We further assume that $p(x,\theta)$ is an exponential family of probability densities \cite{Amari01}:
\begin{equation}
p(x,\theta) = \exp[\theta^Tc(x) - \psi(\theta)], 
\label{eq:expfamily}
\end{equation}
where $\theta = (\theta_1,\ldots,\theta_r)^T$ is the natural parameter, 
$c(x) = (c_1(x), \ldots,c_r(x))^T$ is the sufficient statistic for $\theta$ and $\exp[-\psi(\theta)]$ is the normalization factor.  
Substituting Eq.~(\ref{eq:expfamily}) into Eq.~(\ref{eq:projectionFP}) leads to 
the projection approximation onto the exponential family:
\begin{equation}
\frac{d\theta}{dt} = g^{-1}(\theta)\mathrm{E}[\mathcal{L}c],
\label{eq:expfilter}
\end{equation}
where $\mathcal{L}$ is the backward diffusion operator:
\begin{equation}
\fl
\mathcal{L} = \sum_{i=1}^n \Bigg[\sum_{k=1}^m\Delta_{ik}h_k(x)\Bigg] \frac{\partial}{\partial x_i} + \frac{1}{2}
\sum_{i,j=1}^n
\Bigg[\sum_{k=1}^m \Delta x_{ik}h_k(x)\Delta x_{jk}\Bigg]
\frac{\partial^2}{\partial x_i \partial x_j}.
\label{eq:backward}
\end{equation}

%%%%%%%%%%%%%%%%%%%%%%%%%%%%%%%%%%
\subsubsection{Bayesian update}

Let $\theta(t_i)$ be the solution of Eq.~(\ref{eq:expfilter}) at time $t_i$. 
At time $t_i$, the observation $y_i$ is combined with $p(x,\theta(t_i))$ through Bayes' rule, leading to the posterior probability density of $x$:
\begin{equation}
p^+(x,t_i) = \frac{p(y_i|x)p(x,\theta(t_i))}{\int p(y_i|x)p(x,\theta(t_i))dx},
\label{eq:Bayes}
\end{equation}
where $p(y_i|x)$ is the likelihood function of the observation model (\ref{eq:observation}). 
If $p(x,\theta)$ is a conjugate family for $p(y_i|x)$, 
then the posterior probability density is in the same exponential family (\ref{eq:expfamily}):
\begin{equation}
p^+(x,t_i) = \exp[\theta^+(t_i)^Tc(x)- \phi(\theta^+(t_i))],
\end{equation}
where $\theta^+(t_i)$ is the parameter updated by Bayes' rule. 

The filtering algorithm is summarized in the following two steps: 
\begin{enumerate}
\item (Prediction step)
Solve the ODE (\ref{eq:expfilter}) from time $t_{i-1}$ to $t_i$ with initial conditions $\theta^+(t_{i-1})$ to obtain $\theta(t_i)$. 
\item (Correction step)
Update the parameter $\theta(t_i)$ to $\theta^+(t_i)$ by Bayes' rule (\ref{eq:Bayes}).
\end{enumerate}
Filtering is performed by executing these two steps recursively from time $t_1$ to $t_N$.

%%%%%%%%%%%%%%%%%%%%%%%%%%%%%%%%%%
\subsection{Choice of probability distributions}
%%%%%%%%%%%%%%%%%%%%%%%%%%%%%%%%%%

We use two specific probability distributions for $p(x,\theta)$ to illustrate our method.

%%%%%%%%%%%%%%%%%%%%%%%%%%%%%%%%%
\subsubsection{Gaussian distribution} \label{sec:Gaussian}

Consider a multi-dimensional Gaussian distribution with mean vector $\mu$ and covariance matrix $Q$:
\begin{equation}
p(x, \mu,Q) = 
(2\pi)^{-n/2}|Q|^{-1/2}\exp \left[ -\frac{1}{2}(x-\mu)^TQ^{-1}(x-\mu) \right]. 
\label{eq:gausspdf}
\end{equation}
It is easily confirmed that the Gaussian distribution belongs to the exponential families (\ref{eq:expfamily}).
The projection approximation (\ref{eq:expfilter}) is obtained as (see \ref{appendix:GP})
\begin{eqnarray}
\frac{d\mu}{dt} &=& A \mathrm{E}[h(x)],  \label{eq:GP_mean} \\
\frac{dQ}{dt} &=& Q\mathrm{E}[J_h(x)]^TA^T + A\mathrm{E}[J_h(x)]Q +
A\mathrm{E}[H(x)]A^T, \label{eq:GP_cov}
\end{eqnarray}
where
\begin{equation}
J_h(x) := \frac{\partial h(x)}{\partial x}
\end{equation}
is the Jacobian matrix of $h(x)$.
Note that Eqs.~(\ref{eq:GP_mean})-(\ref{eq:GP_cov}) are expressed with $(\mu,Q)$ instead of the natural parameter $\theta$ of the exponential family.
For systems with reactions of order three or higher, 
$h(x)$ contains polynomials in the variables of order three or higher, so that  
Eqs.~(\ref{eq:GP_mean})-(\ref{eq:GP_cov}) depend on moments of order three or larger;
these moments can be computed with $\mu$ and $Q$ due to the Gaussian assumption, and therefore Eqs.~(\ref{eq:GP_mean}) and (\ref{eq:GP_cov}) are closed for such systems. 
We also point out that 
the Gaussian projection is equivalent to the normal moment-closure approximation (see \ref{appendix:normal_moment_closure} for proof). 

Since both $p(x,\theta(t_i))$ and $p(y_i|x)$ in Eq.~(\ref{eq:Bayes}) are Gaussian distributions, $p^+(x, t_i)$ is also Gaussian, and its mean vector $\mu^+(t_i)$ and covariance matrix $Q^+(t_i)$ are computed using the standard Kalman filter recursion as 
\begin{eqnarray}
\mu^+(t_i) &=& \mu(t_i) + K_i\{y_i - G\mu(t_i)\}, \label{eq:post_mean} \\ 
Q^+(t_i) &=& Q(t_i) - K_iGQ(t_i), \label{eq:post_cov}
\end{eqnarray}
where 
\begin{equation}
K_i = Q(t_i)G^T\{GQ(t_i)G^T+V\}^{-1}
\label{eq:post_gain}
\end{equation}
is the Kalman gain \cite{Sarkka13}.

%%%%%%%%%%%%%%%%%%%%%%%
\subsubsection{Quartic polynomial} \label{sec:quartic}

Another example is an exponential family of probability distributions with quartic polynomials in the exponent: 
$c(x) = (x,x^2,x^3,x^4)^T$ $(x\in\mathbb{R}^1)$ and $\theta = (\theta_1,\theta_2,\theta_3,\theta_4)^T$. 
A characteristic of this exponential family is that it allows bimodality. 
We briefly summarize how to compute the Fisher information matrix $g(\theta)$ and the moments $\eta_i:=\mathrm{E}[x^i]$ $(i=1,2,\ldots)$ that are required to solve the ODE (\ref{eq:expfilter}) (see \cite{Brigo99} for details).
\begin{enumerate}
\item %%% 1 %%%
For $i=0,1,2$, compute the following integral numerically:
\begin{equation}
I_i(\theta) = \int_{-\infty}^{\infty}x^i\exp(\theta_1x+\theta_2x^2+\theta_3x^3+\theta_4x^4)dx
\end{equation}
and $\eta_i = I_i(\theta)/I_0(\theta)$.
\item %%% 2 %%%
Compute recursively the higher-order moments $\eta_i(\theta)$, $i\ge3$ by 
\begin{equation}
\fl
\eta_i(\theta) = -\frac{1}{4\theta_4}\{ (i-3)\eta_{i-4}(\theta) + \theta_1\eta_{i-3}(\theta) 
+ 2\theta_2\eta_{i-2}(\theta) + 3\theta_3\eta_{i-1}(\theta) \}.
\end{equation}
\item %%% 3 %%%
Compute the Fisher information matrix $g(\theta) = (g_{ij}(\theta))$ where 
\begin{equation}
g_{ij}(\theta) = \eta_{i+j}(\theta) - \eta_i(\theta)\eta_j(\theta).
\end{equation}
\end{enumerate}
For this exponential family distribution, the parameter update through Bayes' rule (\ref{eq:Bayes}) becomes 
\begin{eqnarray}
\left( \begin{array}{c}
\theta_1^+(t_i) \\
\theta_2^+(t_i) \\
\theta_3^+(t_i) \\ 
\theta_4^+(t_i) 
\end{array} \right)
=
\left( \begin{array}{c}
\theta_1(t_i) + \frac{Gy_i}{V} \\
\theta_2(t_i) - \frac{G^2}{2V} \\
\theta_3(t_i) \\
\theta_4(t_i)
\end{array} \right).
\end{eqnarray}

%%%%%%%%%%%%%%%%%%%%%%%%%%%%%%%%%%%%%%%%%%%
\section{Results}
%%%%%%%%%%%%%%%%%%%%%%%%%%%%%%%%%%%%%%%%%%%

We illustrate our method on two reaction networks, and compare it to an approximate filter based on the LNA in their filtering performances. 
The LNA-based filter is briefly summarized in \ref{appendix:LNA}.
Hereafter, we label the projection-based filter onto Gaussian distributions ``GPF" and that onto quartic polynomial exponential distributions ``QPF". 

%%%%%%%%%%%%%%%%%%%%%%%%%%%%%%%%%%%%%%%%%%%
\subsection{Bistable system}

\begin{figure}[t]
\begin{center}
\includegraphics[width=15cm]{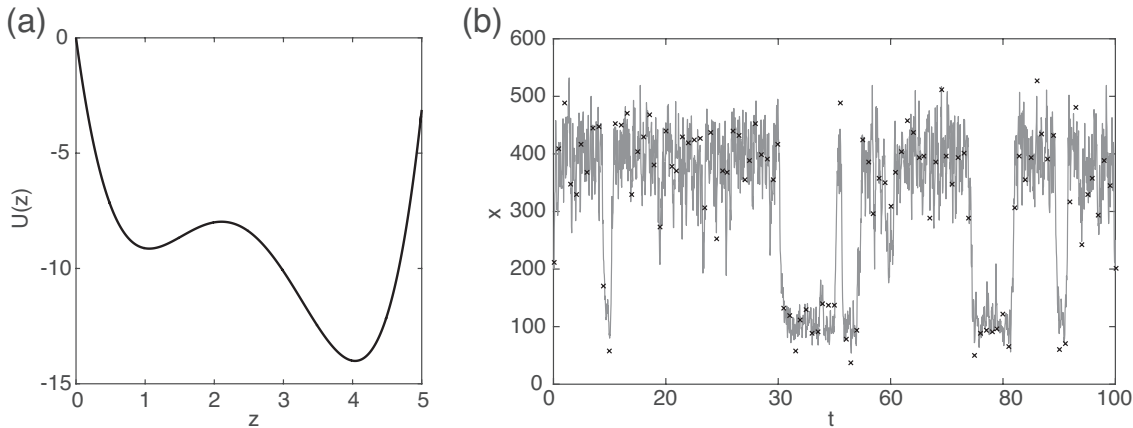}
\end{center}
\caption{
(a) Potential $U(z)$ has two local minima at $z=1.06$ and $z=4.04$. 
(b) Gray line represents a sample path of $x(t)$ simulated with the stochastic simulation algorithm, and crosses represent observations with the noise variance $V=500$.
}
\label{fig:bistable}
\end{figure}

We first consider the following reaction network consisting of a single species \cite{Erban09}:
\begin{eqnarray}
\emptyset \autorightleftharpoons{$k_1$}{$k_2$} X, \quad
2X \autorightleftharpoons{$k_3$}{$k_4$} 3X. \nonumber
\end{eqnarray}
The net effect matrix and the reaction rate vector, respectively, are given by 
 \begin{equation}
A = (1, -1, 1 ,-1),
\label{eq:neteffect_bistable}
\end{equation}
and 
\begin{equation}
h(x) = (k_1, k_2x, k_3x(x-1), k_4x(x-1)(x-2))^T.
\label{eq:propensity_bistable}
\end{equation}
The rate equation (\ref{eq:deterministic}) for $z\Omega\gg1$ is given by 
\begin{equation}
\frac{dz}{dt} = -\frac{dU(z)}{dz},
\end{equation}
where 
$U(z)$ is the potential:
\begin{equation}
U(z) = -\tilde{k}_1z + \frac{\tilde{k}_2}{2}z^2 - \frac{\tilde{k}_3}{3}z^3 + \frac{\tilde{k}_4}{4}z^4,
\label{eq:potential}
\end{equation}
with the rescaled rate constants:
\begin{equation}
\tilde{k}_1 = \frac{k_1}{\Omega}, \ 
\tilde{k}_2 = k_2, \ 
\tilde{k}_3 = \Omega k_3, \ 
\tilde{k}_4 = \Omega^2k_4.
\end{equation}
The parameter values were considered to be 
$\tilde{k}_1 = 22.5$, $\tilde{k}_2 = 37.5$, $\tilde{k}_3 = 18$ and $\tilde{k}_4 = 2.5$, 
with which the potential (\ref{eq:potential}) has two local minima (Figure~\ref{fig:bistable}a). 
The stochastic version of the reaction network with $\Omega=100$ was simulated using the SSA. 
A sample path is shown in Figure~\ref{fig:bistable}b (gray line) wherein we see that the reaction network exhibits stochastic switching between the two states that correspond to the two local minima of the potential.

For this reaction network, we applied the GPF, QPF and LNA. 
A numerical study was conducted using the following steps: 
First, the reaction network was simulated with the SSA in a time interval $T=100$ to generate a sample path, $\{x(t), 0\le t \le T\}$ (Figure~\ref{fig:bistable}b, gray line).
The observations, $\{y_i, i=1,\ldots,N\}$, were simulated using Eq.~(\ref{eq:observation}), 
where we set $G=1$.
The inter-observation interval,
$\Delta:=t_i-t_{i-1}$, ranged from $0.1$ to $1$, 
and the variance of the observation noise, $V$, ranged from $500$ to $5,000$ 
(Figure~\ref{fig:bistable}b; crosses represent the observations with $\Delta=1$ and $V=500$). 
The three approximate filters were then performed to estimate the simulated path from the observations.

\begin{figure}[tb]
\begin{center}
\includegraphics{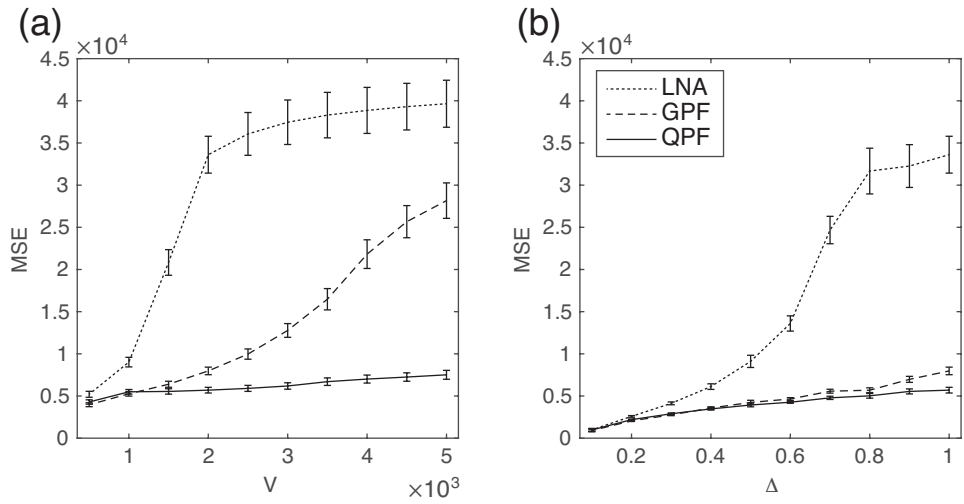}
\end{center}
\caption{
Mean squared error (MSE) between true and estimated paths (a) as a function of noise variance $V$ with $\Delta=1$ and (b) as a function of interval $\Delta$ with $V=2,000$.
Solid, dashed and dotted lines represent MSE for QPF, GPF and LNA, respectively. 
Mean squared errors at each point were calculated with 20 repetitions.  
MSE for QPF is smaller than that for LNA and GPF over the range of $V$ and $\Delta$. 
}
\label{fig:bistable_mse}
\end{figure}

\begin{figure}[tb]
\begin{center}
\includegraphics{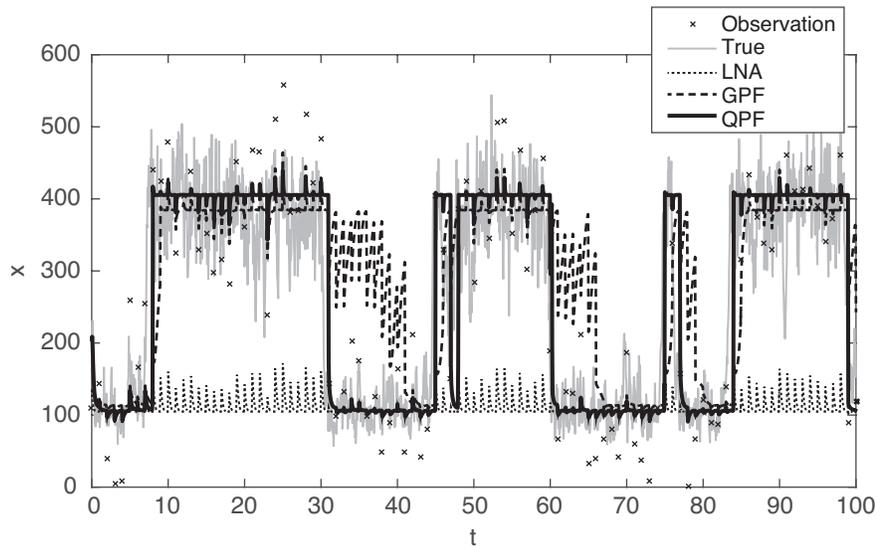}
\end{center}
\caption{
Sample simulated paths for $V=3,000$ and $\Delta=1$. 
Gray line represents true path, and solid, dashed and dotted lines represent paths estimated by QPF, GPF and LNA, respectively.  
While QPF captures the abrupt jumps, GPF and LNA fail, resulting in the large estimation error. 
}
\label{fig:bistable_est}
\end{figure}

To quantify the extent to which the approximate filters estimate the true path, we employed a {\it maximum a posteriori} (MAP) estimate, $\hat{x}(t)$, for each filter, and computed the mean squared error (MSE) between the true and estimated paths: 
\begin{equation}
\mathrm{MSE} = \frac{1}{T}\int_0^T | x(t)-\hat{x}(t) |^2dt.
\end{equation}
We plotted the MSE for the three approximate filters as a function of $V$ (Figure~\ref{fig:bistable_mse}a) and as a function of $\Delta$ (Figure~\ref{fig:bistable_mse}b). 
The difference in the MSE among the three filters is small when $V$ or $\Delta$ is small. 
The MSE for the LNA increases more than that for the GPF and QPF as $V$ or $\Delta$ is increased. 
In particular, the MSE for the QPF remains relatively small over the range of $V$ and $\Delta$. 
Figure \ref{fig:bistable_est} depicts sample paths estimated by the three filters for $V=3,000$ and $\Delta=1$;  as seen in this figure, while the QPF can capture the sharp transitions from one local equilibrium state to the other, the GPF and LNA fail, resulting in the large estimation error.
These results suggest that for the reaction network with bistability, the QPF performs better that the GPF and LNA; 
the superiority of the QPF over the others stands out for noisy and sparse observations.

%%%%%%%%%%%%%%%%%%%%%%%%%%%%%%%%%%%%%%%
\subsection{Reaction network with limit cycle}

\begin{figure}[tb]
\begin{center}
\includegraphics{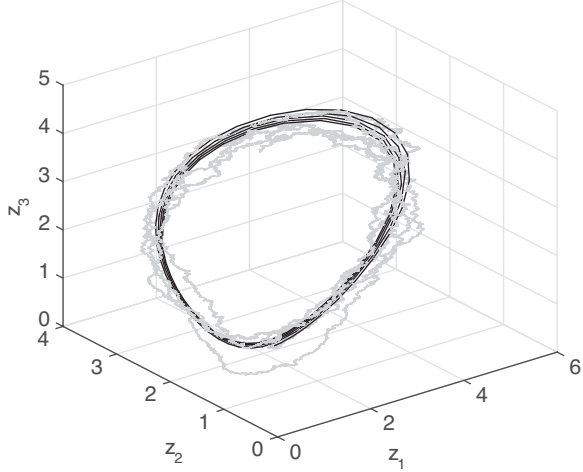}
\end{center}
\caption{
The phase space $(z_1, z_2, z_3)$. 
Black line represents a solution of ordinary differential equations (\ref{eq:limcyc_rate_1})-(\ref{eq:limcyc_rate_3}), 
and gray line represents a sample path of stochastic model. 
}
\label{fig:limcyc}
\end{figure}

Next, we consider a reaction network consisting of three species, $X = (X_1,X_2,X_3)$, which follow a set of five reactions \cite{Wilhelm95}:
\begin{eqnarray}
X_1 \autorightarrow{$k_1$}{} 2X_1, \quad 
X_1 + X_2  \autorightarrow{$k_2$}{} X_2, \quad
X_2  \autorightarrow{$k_3$}{} \emptyset, \nonumber\\
X_1  \autorightarrow{$k_4$}{} X_3, \quad
X_3 \autorightarrow{$k_5$}{} X_2. \nonumber
\end{eqnarray}
The net effect matrix and the reaction rate vector, respectively, are given by 
\begin{equation}
A = \left(
\begin{array}{rrrrrrr}
1 & -1  &  0 & -1 & 0  \\
0 &  0  & -1 &  0 &  1  \\
0 &  0  &  0 &  1 & -1 \\
\end{array}
\right),
\end{equation}
\begin{equation}
h(x) = (k_1x_1, k_2x_1x_2, k_3x_2, k_4x_1, k_5x_3)^T.
\end{equation}
The rate equation (\ref{eq:deterministic}) is derived as 
\begin{eqnarray}
\frac{dz_1}{dt} = (\tilde{k}_1 - \tilde{k}_4)z_1 - \tilde{k}_2z_1z_2, \label{eq:limcyc_rate_1} \\
\frac{dz_2}{dt} =  -\tilde{k}_3z_2 + \tilde{k}_5z_3, \label{eq:limcyc_rate_2} \\
\frac{dz_3}{dt} =  \tilde{k}_4z_1 - \tilde{k}_5z_3, \label{eq:limcyc_rate_3}
\end{eqnarray}
where the reaction rate constants are rescaled as 
\begin{equation}
\tilde{k}_1 = k_1, \ 
\tilde{k}_2 = \Omega k_2, \ 
\tilde{k}_3 = k_3, \ 
\tilde{k}_4 = k_4, \
\tilde{k}_5 = k_5.
\end{equation}
The values of the rate constants were chosen as $\tilde{k}_1 = 3.1$, $\tilde{k}_2 = 1$, $\tilde{k}_3 = 1$, $\tilde{k}_4 = 1$ and $\tilde{k}_5 = 1$.
Figure~\ref{fig:limcyc} depicts the phase space $(z_1,z_2,z_3)$ wherein an illustrative path of the rate equation is plotted (black line), showing that it converges to the limit cycle.
The stochastic version of the reaction network with $\Omega=100$ was simulated with the SSA. 
A sample path of the rescaled variable $x/\Omega$ was also plotted in Figure~\ref{fig:limcyc} (gray line).

\begin{figure}[tb]
\begin{center}
\includegraphics{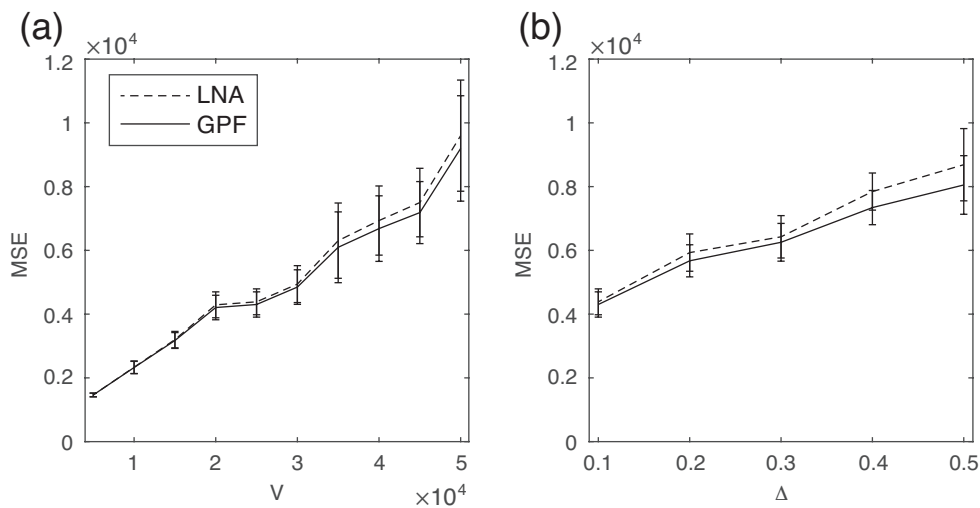}
\end{center}
\caption{
Mean squared error (MSE) between true and estimated paths (a) as a function of noise variance $V$ with $\Delta=0.1$ and (b) as a function of interval $\Delta$ with $V=2,500$ for GPF (solid line) and for LNA (dashed line). 
Mean squared errors at each point were calculated with 20 repetitions.  
MSE for GPF is slightly smaller than that for LNA. 
}
\label{fig:limcyc_mse}
\end{figure}

We applied the GPF and the LNA for this reaction network. 
A numerical study for this reaction network was performed using the same procedure as for the bistable system.  
The duration of the simulation interval was chosen as $T=30$. 
The parameter of the observation model (\ref{eq:observation}) was considered to be $G=(1,0,0)$. 
The inter-observation interval, $\Delta:=t_i-t_{i-1}$, ranged from $0.1$ to $0.5$, 
and the variance of the observation noise, $V$, ranged from $5,000$ to $50,000$.
We plotted the MSE between the true and estimated paths as a function of $V$ (Figure~\ref{fig:limcyc_mse}a) and as a function of $\Delta$ (Figure~\ref{fig:limcyc_mse}b) for the GPF (solid line) and for the LNA (dashed line). 
We see that the MSE for the GPF is smaller than that for the LNA. 
However, a very little difference in the MSE between these two methods is observed.

%%%%%%%%%%%%%%%%%%%%%%%%%%%%%%%%%%%%%%%%%%%
\section{Discussion}
%%%%%%%%%%%%%%%%%%%%%%%%%%%%%%%%%%%%%%%%%%%

In this section, we compared between the projection and moment-closure approximations. 
As seen in the section~\ref{sec:Gaussian} and \ref{appendix:normal_moment_closure}, the projection approximation onto Gaussian distributions is equivalent to the moment-closure approximation based on the same Gaussian distributions. 
However, the projection approximation does not always coincide with moment-closure approximations even if these share a common probability distribution. 
A difference between the two approximation techniques is that while moment-closures yield ODEs for the moments $\mathrm{E}(x^i)$, 
the projection method produces ODEs for the natural parameter $\theta$ of exponential family distributions, which is related to the expectation of the sufficient statistic $c(x)$ \cite{Amari01}.

We illustrate this difference using a reaction network consisting of single species and at most bimolecular reactions:
\begin{equation}
A=(a_1, a_2), \quad h(x)=(k_1x, k_2x(x-1))^T,
\end{equation}
and using gamma distributions for the base probability distributions. 
The probability density of a gamma distribution is given by 
\begin{equation}
p(x,\mu,\kappa) = \frac{\kappa^{\kappa}x^{\kappa-1}}{\mu^{\kappa}\Gamma(\kappa)}e^{-\frac{\kappa x}{\mu}},
\label{eq:gammapdf}
\end{equation}
whose mean and variance are $\mathrm{E}(x)=\mu$ and $\mathrm{Var}(x)=\mu^2/\kappa$, respectively. 
Eq.~(\ref{eq:gammapdf}) can be rewritten in the form (\ref{eq:expfamily}) with the natural parameter $\theta = (-\kappa/\mu, \kappa-1)$ and the sufficient statistic $c(x) = (x, \log x)$. 
The expectations of $c(x)$ is expressed with $(\mu,\kappa)$ as 
\begin{equation}
\mathrm{E}[c(x)] 
= 
(\mathrm{E}[x], \mathrm{E}[\log x])
=
(\mu,  \varphi(\kappa) - \log\kappa + \log\mu),
\end{equation}
where $\varphi(\kappa):=\frac{d}{d\kappa}\log\Gamma(\kappa)$ is the digamma function.
The Fisher information matrix of the gamma distribution with respect to $(\mu,\kappa)$ is given by 
\begin{equation}
g(\mu,\kappa) = 
\left(\begin{array}{cc}
\kappa/\mu^2 & 0 \\
0 & \dot{\varphi}(\kappa) - \kappa^{-1}
\end{array}\right).
\end{equation}
Using these quantities, the projection approximation of the reaction network onto the gamma distributions is derived as
\begin{eqnarray}
\fl
\frac{d\mu}{dt} = (a_1k_1-a_2k_2)\mu + a_2k_2\mu^2 + \frac{a_2k_2\mu^2}{\kappa}, 
\label{eq:pa_gamma_mu} \\
\fl
\frac{d\kappa}{dt} = \frac{1}{1-\kappa\dot{\varphi}(\kappa)}
\Bigg\{ a_2k_2\mu  + \frac{a_2^2k_2\kappa}{2} + \frac{(a_1^2k_1-a_2^2k_2)\kappa^2}{2\mu(\kappa-1)} \Bigg\}. \label{eq:pa_gamma_kp}
\end{eqnarray}
On the other hand, the moment-closure approximation based on the gamma distributions yields a set of ODEs for $\mu$ and $\sigma^2:=\mathrm{Var}(x)$:
\begin{eqnarray}
\fl
\frac{d\mu}{dt} = (a_1k_1-a_2k_2)\mu + a_2k_2\mu^2 + a_2k_2\sigma^2, 
\label{eq:mc_gamma_mu} \\
\fl
\frac{d\sigma^2}{dt} = 2(a_1k_1-a_2k_2)\sigma^2 + \frac{4a_2k_2(\sigma^2+\mu^2)\sigma^2}{\mu} + (a_1^2k_1-a_2^2k_2)\mu + a_2^2k_2(\sigma^2+\mu^2), \label{eq:mc_gamma_var}
\end{eqnarray}
where we used $\mathrm{E}(x^3)=(\mu^2+2\sigma^2)(\mu^2+\sigma^2)/\mu$ to derive Eq.~(\ref{eq:mc_gamma_var}).

%%%%%%%%%%%%%%%%%%%%%%%%%%%%%%%%%%%%%%%%%%%
\section{Conclusion}
%%%%%%%%%%%%%%%%%%%%%%%%%%%%%%%%%%%%%%%%%%%

This study concerned the filtering problem for stochastic reaction networks. 
The difficulty in deriving filtering algorithms stems from the analytical intractability of the master equation. We applied the projection method to derive approximate filters. 

The projection method provides a flexible framework for approximating reaction networks, as any probability distribution in exponential families fits this method. 
We demonstrated it on the two reaction networks. 
In particular, the projection-based filter with quartic polynomials exhibited much better performance than the other methods for the reaction system with bistability (Figure~\ref{fig:bistable_mse}), due to its capability to accommodate bimodal distributions. 

We note that numerical methods based on particle filtering have been proposed for the inference of reaction networks \cite{6161329}, 
which would be applicable for the considered molecule numbers. 
It would be interesting to compare the projection-based filter with these methods in terms of the balance between accuracy and computational time of estimation.  

We considered the filtering problem wherein the objective is to estimate the state paths from the observations obtained up to the current time; 
another related problem is {\it smoothing}, which aims to estimate the state paths from the whole observations \cite{Sarkka13,Anderson72}. 
The smoothing equation is not analytically tractable except in the case of linear Gaussian systems, hence approximate methods must be developed along the same line.

It is also an important issue to infer the model parameters \cite{Golightly06,Golightly11}. 
Methods for estimating the reaction rate constants have been developed 
using the LNA, the system-size expansion and moment-closure approximations \cite{Milner13,Frohlich16,Komorowski09,Finkenstadt13,Ruttor09,Stathopoulos13}.  
In addition, 
it is difficult to distinguish between process and measurement noise; the simultaneous estimation of the noise parameters would render the problem substantially more challenging. We leave it for future research.

%%%%%%%%%%%%%%%%%%%%%%%%%%%%%%%%%%%%%%%%%%%
\section*{Acknowledgments}
%%%%%%%%%%%%%%%%%%%%%%%%%%%%%%%%%%%%%%%%%%%

The author would like to thank Prof. Shinomoto for valuable comments.
The author would also like to thank the reviewers for their comments that help improve the manuscript.

%%%%%%%%%%%%%%%%%%%%%%%%%%%%%%%%%%%%%%%%%%%
\appendix
\section{Derivation of the Gaussian projection} \label{appendix:GP}
%%%%%%%%%%%%%%%%%%%%%%%%%%%%%%%%%%%%%%%%%%%

The probability density of the multi-dimensional Gaussian distribution (\ref{eq:gausspdf}) is rewritten in the form of (\ref{eq:expfamily}) with
\begin{equation}
\psi(\theta) = \frac{1}{2} (\mu^TQ^{-1}\mu + n\log 2\pi + \log|Q|),
\end{equation}
\begin{equation}
c(x) =
\left(\begin{array}{c}
x \\  \mathrm{col}(xx^T)
\end{array}\right),
\label{eq:suffstat}
\end{equation}
and 
\begin{equation}
\theta^T 
=
(\theta_1,\theta_2,\ldots,\theta_n, \mathrm{col}(\Phi)^T),
\end{equation}
where $\Phi =(\phi_{ij}) := -\frac{1}{2}Q^{-1}$ and 
\begin{equation}
\theta_i = -\sum_{j=1}^n(\phi_{ij}+\phi_{ji})\mu_j, \quad i=1,\ldots,n.
\end{equation}
Here, for a $n\times n$ matrix $B$ we defined the {\it column} operation as
\begin{equation}
\mathrm{col}(B) = 
\left(\begin{array}{c}
B(1) \\ B(2) \\ \vdots \\ B(n)
\end{array}\right),
\end{equation}
where $B(i)$ is the $i$th column of $B$.

We introduce the following two parameterizations: 
\begin{equation}
\zeta = 
\left( \begin{array}{c}
\mu \\ \mathrm{col}(\Phi)
\end{array} \right), 
\quad
\eta = 
\left( \begin{array}{c}
\mu \\ \mathrm{col}(Q)
\end{array} \right),
\end{equation}
and consider the transformations of parameters, $\theta \mapsto \zeta \mapsto \eta$. 
The Jacobian matrices of these transformations, $J_{\theta}(\zeta):=\partial \theta/\partial \zeta$ and $J_{\zeta}(\eta):=\partial \zeta/\partial \eta$, are given by 
\begin{equation}
\label{eq:Jacobians_GP}
J_{\theta}(\zeta) =
\left( \begin{array}{cc}
Q^{-1} &  M^T \\
\mathbf{0}_{n^2\times n} & \mathbf{1}_{n^2}
\end{array} \right), 
\quad
J_{\zeta}(\eta) = 
\left( \begin{array}{cc}
\mathbf{1}_n & \mathbf{0}_{n\times n^2} \\
\mathbf{0}_{n^2\times n} & J_{\Phi}(Q)
\end{array} \right),
\end{equation} 
where 
 $J_{\Phi}(Q):= \partial  \mathrm{col}(\Phi)/\partial \mathrm{col}(Q)$ 
is the Jacobian matrix of $\mathrm{col}(\Phi)$, and 
$M$ is a $n^2\times n$ matrix given by
\begin{equation}
M = -\mu \otimes \mathbf{1}_n - \mathbf{1}_n \otimes \mu,
\end{equation}
where $\otimes$ is the tensor product for two matrices $B=(b_{ij})$ and C defined by  
\begin{equation}
B \otimes C = 
\left( \begin{array}{cccc}
b_{11}C & b_{12}C & \cdots & b_{1n}C \\
b_{21}C & b_{22}C & \cdots & b_{2n}C \\
\vdots & \vdots & \ddots & \vdots \\
b_{n1}C & b_{n2}C & \cdots & b_{nn}C \\
\end{array}\right).
\end{equation} 
By transforming the parameters as $\theta \mapsto \zeta \mapsto \eta$, we can express Eq.~(\ref{eq:expfilter}) as 
\begin{equation}
\frac{d\eta}{dt} = g^{-1}(\eta)J_{\zeta}(\eta)^TJ_{\theta}(\zeta)^T\mathrm{E}[\mathcal{L}c],
\label{eq:expfilter_eta}
\end{equation}
where $g(\eta)$ is the Fisher information matrix of $\eta$, given by
\begin{equation}
\label{eq:g_eta}
g(\eta) = 
\left( \begin{array}{cc}
Q^{-1} & \mathbf{0}_{n\times n^2} \\
\mathbf{0}_{n^2\times n} & \mathcal{I}(Q)
\end{array} \right).
\end{equation}
In Eq.~(\ref{eq:g_eta}), $\mathcal{I}(Q)$ is the Fisher information matrix of $\mathrm{col}(Q)$, which is expressed by the change of parameter as 
\begin{equation}
\mathcal{I}(Q) = J_{\Phi}(Q)^T\mathcal{I}(\Phi)J_{\Phi}(Q).
\end{equation}
Since $\mathrm{col}({\Phi})$ is the natural parameter of the Gaussian distribution (\ref{eq:gausspdf}), and $\mathrm{col}({Q})$ is the corresponding expectation parameter, 
$\mathcal{I}(\Phi)$ is given by the Jacobian matrix $\partial \mathrm{col}(Q)/\partial \mathrm{col}(\Phi) = J_{\Phi}^{-1}(Q)$ \cite{Amari01}. 
Thus, we obtain 
\begin{equation}
\label{eq:I_Q}
\mathcal{I}(Q) = J_{\Phi}(Q)^T. 
\end{equation}
The factor $\mathrm{E}[\mathcal{L}c]$ in Eq.~(\ref{eq:expfilter_eta}) is obtained from Eqs.~(\ref{eq:backward}) and (\ref{eq:suffstat}) as
\begin{equation}
\label{eq:EL_c}
\fl
\mathrm{E}[\mathcal{L}c] = 
\left(\begin{array}{cc}
 A\mathrm{E}[h(x)] \\
 \mathrm{col}\{ A\mathrm{E}[h(x)x^T] + \mathrm{E}[xh(x)^T]A^T + A\mathrm{E}[H(x)]A^T \}
\end{array}\right).
\end{equation}
Substituting Eqs.~(\ref{eq:Jacobians_GP}), (\ref{eq:g_eta}), (\ref{eq:I_Q}) and (\ref{eq:EL_c}) into Eq.~(\ref{eq:expfilter_eta}) leads to 
\begin{eqnarray}
\fl
\frac{d\eta}{dt} 
=
\left(\begin{array}{c}
A\mathrm{E}[h(x)] \\
MA\mathrm{E}[h(x)] + 
 \mathrm{col}\{ A\mathrm{E}[h(x)x^T] + \mathrm{E}[xh(x)^T]A^T + A\mathrm{E}[H(x)]A^T \}
\end{array}\right).
\label{eq:GP_A1}
\end{eqnarray}
Using the following equality,
\begin{eqnarray}
MA\mathrm{E}[h(x)] 
&=& 
-(\mu \otimes \mathbf{1}_{n})A\mathrm{E}[h(x)] - (\mathbf{1}_{n} \otimes \mu) A\mathrm{E}[h(x)] \nonumber\\
&=&
-\mathrm{col}\{A\mathrm{E}[h(x)]\mu^T)\} - \mathrm{col}\{\mu\mathrm{E}[h(x)]^TA^T\}, 
\end{eqnarray}
the second row of Eq.~(\ref{eq:GP_A1}) can be rewritten as 
\begin{eqnarray}
\fl
\lefteqn{
\mathrm{col}\{
A\mathrm{E}[h(x)(x-\mu)^T] + \mathrm{E}[(x-\mu)h(x)^T]A^T + A\mathrm{E}[H(x)]A^T \}
}\hspace{1cm}\nonumber\\
&=&
\mathrm{col}\{ A\mathrm{E}[J_h(x)]Q + Q\mathrm{E}[J_h(x)]^TA^T + A\mathrm{E}[H(x)]A^T \},
\label{eq:GP_A2}
\end{eqnarray}
where the equality follows from the Gaussian assumption. 
From Eqs.~(\ref{eq:GP_A1}) and (\ref{eq:GP_A2}), we obtain the Gaussian projection (\ref{eq:GP_mean})-(\ref{eq:GP_cov}).

%%%%%%%%%%%%%%%%%%%%%%%%%%%%%%%%%%%%%%%%%%%
\section{Derivation of the normal moment-closure approximation} 
\label{appendix:normal_moment_closure}
%%%%%%%%%%%%%%%%%%%%%%%%%%%%%%%%%%%%%%%%%%%

In this appendix, we derive the normal moment-closure approximation for the stochastic reaction networks \cite{Milner13,Goodman53,Gomez-Uribe07,Cseke15}, and show that it is equivalent to the Gaussian projection approximation. 

The mean of $x$ is defined by $\mu = \sum_{x=0}^{\infty}P(x,t)x$, 
where $\sum_{x=0}^{\infty} := \sum_{x_1=0}^{\infty} \sum_{x_2=0}^{\infty}\cdots \sum_{x_n=0}^{\infty}$. 
Then, from Eq.~(\ref{eq:stochastic}) we obtain 
\begin{equation}
\fl
 \frac{d\mu}{dt}
 = 
  \sum_{x=0}^{\infty} \sum_{j=1}^m h_j(x-\Delta x_{\cdot j})P(x-\Delta x_{\cdot j},t) x
-   \sum_{x=0}^{\infty} \sum_{j=1}^m h_j(x)P(x,t)x.
\label{eq:mc_mean_1}
\end{equation}
For each $j=1,\ldots,m$, it follows that 
\begin{eqnarray}
\lefteqn{
\sum_{x=0}^{\infty}h_j(x-\Delta x_{\cdot j})P(x-\Delta x_{\cdot j},t) x 
}\hspace{1cm}\nonumber\\
&=&
\sum_{y=0}^{\infty}h_j(y)P(y,t)(y+\Delta x_{\cdot j}) \nonumber\\
&=&
\sum_{x=0}^{\infty}h_j(x)P(x,t)x + \sum_{x=0}^{\infty}h_j(x)P(x,t)\Delta x_{\cdot j}, 
\label{eq:mc_mean_2}
\end{eqnarray}
where we used the fact that $P(x,t)=0$ and $h(x)=0$ if there exists $i\in\{1,\ldots,n\}$ such that $x_i<0$.
Putting Eq.~(\ref{eq:mc_mean_2}) back into  Eq.~(\ref{eq:mc_mean_1}) leads to 
\begin{eqnarray}
\frac{d\mu}{dt} 
=  
\sum_{j=1}^m \sum_{x=0}^{\infty} h_j(x)P(x,t)\Delta x_{\cdot j}
=
A\mathrm{E}[h(x)]. 
\label{eq:mc_mean}
\end{eqnarray}

%%%%
Next, we consider the second moment, $\mathrm{E}(xx^T) = \sum_{x=0}^{\infty}P(x,t)xx^T$. 
From Eq.~(\ref{eq:stochastic}), the equation for the second moment reads
\begin{equation}
\fl
 \frac{d\mathrm{E}(xx^T)}{dt}
 = 
  \sum_{x=0}^{\infty} \sum_{j=1}^m h_j(x-\Delta x_{\cdot j})P(x-\Delta x_{\cdot j},t) xx^T
-   \sum_{x=0}^{\infty} \sum_{j=1}^m h_j(x)P(x,t)xx^T.
\label{eq:mc_2nd_1}
\end{equation}
In the same manner as Eq.~(\ref{eq:mc_mean_2}), we obtain
\begin{eqnarray}
\fl
\sum_{x=0}^{\infty}h_j(x-\Delta x_{\cdot j})P(x-\Delta x_{\cdot j},t) xx^T
\nonumber\\
=
\sum_{y=0}^{\infty}h_j(y)P(y,t)(y+ \Delta x_{\cdot j})(y+ \Delta x_{\cdot j})^T 
\nonumber\\
=
\sum_{x=0}^{\infty} h_j(x)P(x,t) 
\{
xx^T + x(\Delta x_{\cdot j})^T + (\Delta x_{\cdot j})x^T + (\Delta x_{\cdot j})(\Delta x_{\cdot j})^T
\}.
\label{eq:mc_2nd_2}
\end{eqnarray}
Substituting Eq.~(\ref{eq:mc_2nd_2}) into Eq.~(\ref{eq:mc_2nd_1}) yields
\begin{eqnarray}
\fl
\frac{d\mathrm{E}(xx^T)}{dt}
&=
\sum_{j=1}^m \sum_{x=0}^{\infty} h_j(x)P(x,t)
\{
x(\Delta x_{\cdot j})^T + (\Delta x_{\cdot j})x^T + (\Delta x_{\cdot j})(\Delta x_{\cdot j})^T
\} \nonumber\\
\fl
&=
\mathrm{E}[xh(x)^T]A^T + A\mathrm{E}[h(x)x^T] + A\mathrm{E}[H(x)]A^T.
\label{eq:mc_2nd}
\end{eqnarray}

Taking the derivative of the covariance of $x$, 
$Q:=\mathrm{Cov}(x) = \mathrm{E}(xx^T) - \mu\mu^T$, 
with respect to $t$, and using Eqs.~(\ref{eq:mc_mean}) and Eq.~(\ref{eq:mc_2nd}) leads to the equation for $Q$ as
\begin{eqnarray}
\fl
\frac{dQ}{dt} 
&= 
\frac{d\mathrm{E}(xx^T)}{dt} - \frac{d\mu}{dt}\mu^T - \mu\frac{d\mu^T}{dt} \nonumber\\
\fl
&=
\mathrm{E}[(x-\mu)h(x)^T]A^T + A\mathrm{E}[h(x)(x-\mu)] + A\mathrm{E}[H(x)]A^T \nonumber\\
\fl
&=
Q\mathrm{E}[J_h(x)]^TA^T + A\mathrm{E}[J_h(x)]Q + A\mathrm{E}[H(x)]A^T,
\label{eq:mc_cov}
\end{eqnarray}
where the last equality follows from the Gaussian assumption. 
Thus, we show that the normal moment-closure approximation, (\ref{eq:mc_mean}) and (\ref{eq:mc_cov}), is equivalent to the Gaussian projection approximation, (\ref{eq:GP_mean}) and (\ref{eq:GP_cov}).

%%%%%%%%%%%%%%%%%%%%%%%%%%%%%%%%%%%%%%%%%%%
\section{Approximate filter based on the LNA} \label{appendix:LNA}
%%%%%%%%%%%%%%%%%%%%%%%%%%%%%%%%%%%%%%%%%%%

In this appendix, we derive an approximate filter based on the LNA. 
The LNA, which is the leading-order term in the system size expansion, is given by a Gaussian process, 
$x(t) \sim \mathcal{N}(\Omega\phi(t) + \sqrt{\Omega}m(t), \Omega \Psi(t))$, 
where $\phi(t)$, $m(t)$ and $\Psi(t)$ are obtained by solving the following ODEs:
\begin{eqnarray}
\frac{d\phi}{dt} &=& A\tilde{h}(\phi), \label{eq:LNA1} \\
\frac{dm}{dt} &=& AJ_{\tilde{h}}(\phi)m, \label{eq:LNA2} \\
\frac{d\Psi}{dt} &=& \Psi J_{\tilde{h}}(\phi)^TA^T + AJ_{\tilde{h}}(\phi)\Psi + A\tilde{H}(\phi)A^T, \label{eq:LNA3}
\end{eqnarray}
with a set of initial conditions, $\phi_0$, $m_0$ and $\Psi_0$ \cite{Kampen92}.
Suppose that in solving Eq.~(\ref{eq:LNA1})-(\ref{eq:LNA3}), the initial distribution of $x$ is given by $\mathcal{N}(\mu_0^*,Q_0^*)$.
Then, we may take an arbitrary $\phi_0$, and set $m_0 = \sqrt{\Omega}(\mu_0^*/\Omega-\phi_0)$ and $\Psi_0=Q_0^*/\Omega$. 
The arbitrariness of initial condition can be resolved by choosing $\phi_0 = \mu_0^*/\Omega$, which makes a relative difference of order $\Omega^{-1/2}$ in $x(t)$.
This initial condition leads to $m(t)=0$ for all $t$ as $m_0=0$, and thus $m(t)$ can be omitted from the LNA.

We can construct an approximate filter by using the above LNA for the prediction step \cite{Fearnhead14}. 
Since the approximate state $x(t_i)$ and the observations $y_i$ follow Gaussian distributions, the correction step can be implemented with the standard Kalman recursions (\ref{eq:post_mean})-(\ref{eq:post_gain}). 
To summarize, the filtering algorithm consists of the following two steps:
\begin{enumerate}
\item (Prediction step)
Solve the ODEs:
\begin{eqnarray}
\fl
\frac{d\mu(t)}{dt} = Ah(\mu(t)), \label{eq:LNA_mean} \\
\fl
\frac{dQ(t)}{dt} = Q(t)J_h(\mu(t))^TA^T + AJ_h(\mu(t))Q(t) +
AH(\mu(t))A^T. \label{eq:LNA_cov}
\end{eqnarray}
from time $t_{i-1}$ to $t_i$ with initial conditions $\mu^+(t_{i-1})$ and $Q^+(t_i)$ to obtain $\mu(t_i)$ and $Q(t_i)$. 
\item (Collection step)
Compute the posterior mean $\mu^+(t_i)$ and covariance matrix $Q^+(t_i)$ at time $t_i$ by Eqs.~(\ref{eq:post_mean})-(\ref{eq:post_gain}).
\end{enumerate}
Eqs.~(\ref{eq:LNA_mean}) and (\ref{eq:LNA_cov}) are obtained by rescaling Eqs.~(\ref{eq:LNA1}) and (\ref{eq:LNA3}) with $\mu(t)=\Omega\phi(t)$ and $Q(t) = \Omega\Psi(t)$.
Notice the difference between the Gaussian projection (\ref{eq:GP_mean})-(\ref{eq:GP_cov}) and LNA (\ref{eq:LNA_mean})-(\ref{eq:LNA_cov}). 
In the Gaussian projection, the expectation of $x(t)$ is taken outside of $h(x(t))$ and $J_h(x(t))$, while it is taken inside of these functions in the LNA.
Hence, these two approximations are equivalent for first-order reactions; 
they differ for second- and higher-order reactions.

%%%%%%%%%%%%%%%%%%%%%%%%%%%%%%%%%%%%%%%%%%%
\section*{References}
%%%%%%%%%%%%%%%%%%%%%%%%%%%%%%%%%%%%%%%%%%%
\bibliography{mybib}

\end{document}